\documentclass[12pt]{iopart}
\usepackage{iopams}

\def\om{\omega}
\def\gm{\gamma}
\def\asb{{\bar \alpha_s}}
\def\as{\alpha_s}
\def\eps{\epsilon}

\begin{document}

\jl{4}

\title{Low x and diffraction: theory}

\author{M McDermott}

\address{Theoretical Physics Group, Department of Physics and Astronomy, 
Brunswick St, University of Manchester, Manchester M13 9PL}

\begin{abstract}

Some outstanding issues in high energy scattering are discussed. 
Particular emphasis is placed on recent developments concerning
the next-to-leading log corrections to the BFKL equation.

\end{abstract}

\section{Introduction}

The working group was concerned with high-energy or small-$x$ phenomena in QCD.
In deep inelastic scattering at moderate $x=Q^2/W^2$ it is customary to think 
in the Breit Frame, in which the proton is moving very fast. 
For large $Q^2$, the short distance partonic valence  
structure of the proton is being probed. 
Striking a single valence parton very violently tends 
to break up the proton, hence the adjective deep-inelastic applied to 
this scattering. At small $x$ (less than about $10^{-3}$) the struck partons 
no longer carry significant amounts of the protons momentum and one may think 
of the photon as probing the radiation field of gluons and $q {\bar q}$-pairs 
surrounding the valence partons. One now has the possibility of striking a 
``wee lump'' of this field without disturbing the integrity of the bound state of the proton.
This is diffraction, and the HERA experiments have observed that indeed 
this possibility occurs about $10\%$ of the time even for $Q^2$ as large as
$800$ GeV$^2$ (for a review of the experimental situation in diffraction see \cite{bcox}) .

At small-x it becomes useful to switch frames to the proton's rest frame. 
The photon fluctuates into a partonic system (a $q {\bar q}$ 
dipole to lowest order in $\alpha_s$) 
a large distance ($d \propto 1/x$) upstream of the stationary proton. 
On the timescale of the interaction with the proton, interactions 
within this partonic system are ``frozen''.
Diffractive scattering corresponds to the case in which there is no net
transfer of colour between the proton and the partonic fluctuation of the photon 
which constitute eigenstates of diffraction. 
The probability of the dipole scattering is directly proportional to its transverse area. 
Hence, for small dipoles one expects small cross sections with the proton,   
since all the colour of the dipole is contained within a small transverse area, 
the colour field of the proton, which contains mainly long-distance fluctuations, 
appears transparent to it: this phenomena has become known 
as {\it colour transparency}.
The longitudinally polarised photon is more inclined to fluctuate in a small symmetric system 
(quark and anti-quark carry roughly the same light-cone momentum fraction of the photon, $z \approx 1/2$)
whereas the transversely polarised photon can split either into a 
large asymmetric  or a small symmetric system. Although the large asymmetric 
system ($z \ll 1/2$) is much less likely, if produced it scatters from the protons 
colour field with a much greater likelihood, so that both types of dipole contribute 
equally (to leading twist) in the diffractive cross section.

As $x \rightarrow 0$ this colour transparency picture must ultimately 
breakdown since the growth of the number of partons must reach the level
where the impulse approximation breaks down and partons in the proton begin to recombine  
with each other. Exactly where in $x$ this transition to a new regime 
happens, at a given $Q^2$, is not known at present (although a recent model has this 
saturation picture built in \cite{wgb}). Once the parton picture has broken
 down it is appropriate to pursue the problem using the classical field language 
(see \cite{mkw} and references therein).

\section{Issues in diffraction}

The major issue in diffraction is that of universality of diffractive phenomena.
Regge factorisation remains the main (imperfect) tool to study this.  
In principle we have four types of diffractive (rapidity gap) ``experiment'' 
to compare with one another: 
1) $\gamma^{*} P \rightarrow X + Y $, 2) $\gamma_{direct} P \rightarrow $ jets + Y, 
3) $\gamma_{resolved} P \rightarrow $ X(jets) + Y 4) $PP \rightarrow $ jets or 
 W + gap (or gaps). In 1-3 $X,Y$ are hadronic systems separated by a rapidity gap.
The H1 QCD analysis of diffractive DIS events \cite{h1diff}, 
which require a gluon-dominated Pomeron to reproduce the positive 
scaling violations observed in 1), appear to do a reasonable job on 2) but overshoot 3) \cite{h1dijet} and fail completely to predict the Tevatron data 4). 
In the latter   a very low fraction of diffractive events are observed  
(low gap survival probability)  and a quark-dominated Pomeron seems to be prefered 
(see \cite{whitmore} and references therein). 
This non-universality is almost certainly connected 
with the secondary interactions of ``passenger'' partons 
in the co-moving systems (different in each case) which act to fill-in the gap of
the primary diffractive scatter. Levin and collaborators have an eikonal 
model of multiple interactions which they have applied to the problem of gap survival probability. 
They find broad agreement with the Tevatron dijets data (see e.g. \cite{levin2}). 
However, it is clear that a systematic study addressing the question of gap survival probability
and the breakdown of universality of diffractive exchanges is urgently needed.

Recently, Donnachie and Landshoff \cite{dl} presented an analysis 
of a wide range of HERA data (including all high energy proton and charm structure function data) 
which seemed to require a second Pomeron with a much larger 
intercept (around 1.4). However, it is not clear how the inclusion of 
this additional Pomeron should, or would, affect the total and 
elastic cross-section fits. 
(it depends on what one assumes for the coupling of this secondary Pomeron 
as $Q^2 \rightarrow 0$). 

We now have very good high precision data on the inclusive and diffractive 
cross sections. However, many models with different dynamical assumptions
produce adequate fits to data. Essentially the reason for this is that it 
is easy to tune the input parameters of a model to fit slowly varying 
function of $x$ and $Q^2$. In principle each model should be able to provide 
predictions for $F_L, F_2^{c {\bar c}}, F_L^{D}$ and $F_2^{D, c {\bar c}}$. 
Of course, if $F_L$ and $F_L^{D}$ were to be measured, for example
by lowering the HERA beam energy, this 
would constrain the possible dynamics considerably. It is encouraging
that there are plans and good prospects of making a good 
measurement $R^{D} = F_L^{D} / F_T^{D} $ at nuclear HERA 
(see \cite{nhera} for a summary of the possibilities of nuclear HERA).

Recently it has become fashionable to present the data on $F_2$ using 
the so-called Caldwell plot (see \cite{bcox}). 
From this plot it is observed that $ dF_2/d \ln (Q^2/Q_0^2)$ plotted as a 
function of $x$ has a maximum around 
$x \approx 10^{-3}$. This has been  presented as evidence for a lack of 
gluons at small-$x$.  However as $x$ decreases the typical $Q^2$ in 
each bin is decreasing and it could be that what we are seeing is 
a breakdown of the parton picture at small $Q^2$. The saturation model of 
Wusthoff and Golec-Biernat \cite{wgb} is designed to take this observed feature 
into account and work is under way to apply the same model
to the diffractive data.

\subsection{Exclusive processes}
  
In order to be sure that it is safe to use perturbative QCD,
it is necessary to isolate those diffractive process in which only 
small dipoles contribute, these exclusive processes are known as hard diffractive, e.g. 
exclusive dijet production, deeply virtual Compton scattering 
or heavy vector meson production. At small-$x$ all such processes are 
governed by the exchange of two gluons in the $t$-channel in a 
colour-singlet configuration. It was recently realised that the 
off-diagonal nature of the amplitude  means that one is 
probing new non-perturbative information about the proton's field 
and it is necessary to replace the gluon density with an off-diagonal 
generalization (which involves 
gluon operators sandwiched between different quantum states)
in expressions for the cross sections of exclusive processes 
(see \cite{ji} for a review and references).
In heavy vector meson production in DIS  
this off-diagonality arises from the need to convert a space-like 
photon virtuality into a time-like vector meson mass.
The evolution equations governing the evolution of these new distributions 
in different regimes, as well as various definitions for them, 
are discussed in detail in \cite{ji}.  

The exclusive electro- and photoproduction of (both heavy and light) vector mesons
is now a mature field both experimentally and theoretically (for a recent review see \cite{critt})
Recent  interesting developments include  
the measurements of the ratios $R_{\phi}/R_{\rho}$ and 
$R_{J/\psi}/R_{\rho}$ versus $t$, at high-$|t|$. These measurements probe
short distance part of the wavefunctions of the vector mesons for the first time. 
It also appears that there is evidence for s-channel helicity 
non-conservation in $\rho$-production (see \cite{bcox,ik}).

\section{Next-to-leading order BFKL}

In this section I discuss some of the recent developments in small-x QCD 
arising from the completion of the next-to-leading corrections to the 
BFKL equation by Fadin and Lipatov \cite{fl} (see also \cite{cc}). There have been many papers this year on this subject. 
I will attempt to summarise some of the issues here.

First of all we recall some results from leading order \cite{bfkl} 
(many aspects of the problem are discussed in the recent textbook by Forshaw 
and Ross  \cite{fr}). The BFKL equation is most safely applied to the 
high-energy scattering of two small systems 
(one usually talks of `onia', a bound state of a 
heavy quark and antiquark) in the Regge limit $s \gg |t|,m^2_{\mbox{onia}}$. 
To the leading-logarithmic accuracy in energy, onia-onia cross sections involve a 
four-point function for a gluon ladder in the t-channel 
convoluted with impact factors for each onia at the bottom and top of the ladder.
By performing the convolution with one of the onia one has 
the unintegrated gluon structure function of this onia, $f(x,k^2_{t})$, convoluted 
with the remaining impact factor. The ladder is calculated in the so-called 
multi-Regge kinematics (MRK), i.e. each cell in the ladder is treated in the Regge
limit, and the longitudinal Sudakov components of the vertical lines are 
strongly ordered.  
These components correspond to the light-cone momentum fractions of the external 
onia carried by the gluon concerned. 
This generates a logarithm in energy for each power in $\alpha_s$. 
The BFKL equation is an integral equation for $f$ which sums up all the 
leading-log parts of all such gluon ladders, i.e. all terms  of order
\begin{equation}
\sum_n  ( \as \ln s/s_0)^n.
\end{equation}

Working to leading-log accuracy does not allow the 
scale of $\as$ or the minimal longitudinal energy scale $s_0$
 to be determined. By taking account all such graphs to 
leading-log accuracy, it is seen that the t-channel gluons Reggeize, 
and one has Reggeon-Reggeon gluon (RRg) vertices in the cross-rungs. 
The kernel of the BFKL equation is essentially democratic in its 
choice of momentum scales: it is {\it conformally invariant}, 
i.e. does not change under $k_t^2 \rightarrow 1/k_t^2$, 
(hereafter I will drop the subscript $t$). 
As a result the typical transverse scales are determined 
only by the external particles and
the distribution in $k^2$ diffuses away from the ends of the ladder. 
This conformal invariance is possible only to leading order in logarithms
since the running of the QCD coupling, which comes in at next to leading-order,  
explicitly breaks this scale invariance.

Because of the conformal invariance, the BFKL equation can be solved by taking 
Mellin moments of f :

\begin{equation}
f(s,k^2) \int_{C_{\om}} \frac {d \omega}{2 \pi i} \int_{C_{\gamma}} 
\frac {d \gamma}{2 \pi i}  \left( \frac{s}{s_0} \right)^{\om} \left(\frac{k^2}{k_0^2} 
\right)^{\gm} f(\om,\gm) \, ,
\end{equation}
\noindent where the Mellin space solution is 
\begin{equation}
f (\om,\gm) \propto \frac{1}{\om - \asb \chi(\gm)}.
\end{equation}
If one expands the $\gm$-integral about the saddle point in the 
kernel, $\gm_s = (1/2,0)$, it contains a pole at 
$\om  = \asb \chi (\gm_s) = 4 \ln 2 \asb $ with $\asb = 3 \as / \pi$.  
This leads to the famous rise in energy of the unintegrated structure function:
\begin{equation}
f (s,k^2) \propto \left( \frac{k^2}{k_0^2} \right)^{\frac{1}{2}} \, 
\left( \frac{s}{s_0} \right)^{4 \ln 2 \, \asb}.
\end{equation}

This corresponds to the lowest energy eigenvalue of a 
Schr\"{o}dinger equation in $k^2$, with $\nu =0$ ($\gm = 1/2 + i \nu$): 
\begin{equation}
H^{(1)} \psi_{\nu} (q) = \eps_{\nu}^{(1)}  \psi_{\nu} (q) 
\end{equation}

\noindent with $\eps_{\nu}^{(1)}$ equal to (minus) the BFKL kernel, $ K(\nu) = 
\asb \chi(\gm) = \asb [ 2 \psi(1) - \psi(1/2 + i \nu) - \psi (1/2 - i\nu)]$. 
The eigenfunctions are given by
\begin{equation}
\psi_{\nu} (k^2) = \frac{(k^2)^{i\nu}}{\sqrt{2 \pi^2 k^2}} \, .
\end{equation}

To next-to-leading log accuracy in energy one needs to sum all 
graphs which contains pieces which are one power down in $\as$, i.e. of order
\begin{equation}
\sum_n  \as ( \as \ln s/s_0)^n.
\end{equation}

\noindent These include all the previous gluon ladders evaluated in the 
quasi-multi-Regge kinematics (QMRK) in which one of the strong orderings is relaxed; 
the original graphs, evaluated in MRK, with running coupling at one loop, 
or NLO corrections to the RRg vertices, or NLO gluon reggeization, or  an 
entirely new set of graphs, with real-quark-emission insertions 
on the horizontal rungs (for more details see \cite{fadin}).

In the Hamiltonian formalism At next-to-leading order the Hamiltonian becomes 
$H = H^{(1)} + H^{(2)}$ where the action of the next-to-leading part, $H^{(2)}$,
on the leading-order eigenfunctions is given by 
\begin{eqnarray}
H^{(2)} \psi_{\nu} (q) = \left[ K_{SI} (\nu) + K_r (\nu,q) \right]  \psi_{\nu} (q) 
\label{nlh} 
\end{eqnarray}

The second term in the right hand side of equation (\ref{nlh}) comes from the 
running of the coupling and explicitly breaks the conformal invariance.
The first term corresponds to all the other corrections, it turns out 
that this piece is scale invariant, like the leading order piece denoted 
$\eps_{\nu}^{(1)}$ and leads to a shift in the the leading order 
eigenvalue which depends on the value of $\asb$ given by
\begin{equation}
\eps_{\nu}^{(1)} \rightarrow \eps_{\nu} = \eps_{\nu}^{(1)} \left( 1 - \frac{\asb C(\nu)}{4} \right)
\end{equation}

\noindent where the function $C(\nu)$ can be found in \cite{fl}.
It turns out that if one concentrates on the saddle point method and expands about $\nu =0$ then the shift of the power of energy is huge and negative for reasonable values of $\asb$:
\begin{eqnarray}
\om_p^{(2)} &=& 4 \ln 2 \asb (1 - 6.56 \asb) \\
	    &=& \om_p^{(1)} (1 - 2.366 \om_p^{(1)} ) \, ,
\label{lc}
\end{eqnarray}
\noindent and considerably reduces the strong rise with energy, $s^{\om_p^{(1)}}$ 
found at leading order. For example, taking $\asb = 0.15$,
a large leading-power $\om_p^{(1)} = 0.4$ is reduced  to $\om_p^{(1+2)} = 0.02$. 

Ross \cite{dross} has pointed out that concentrating on the point 
at $\gamma_s = (1/2,0)$ may be misleading, except for very small values of 
$\asb \leq 0.05 $. For larger values the saddle point on the real-$\gm$ axis is replaced by two saddle points off-axis and that it is necessary to expand about
 these new complex conjugate saddle points to get a more accurate answer. 
This may be achieved by expanding the kernel to order $\Or (\nu^4)$. 
The saddle point method then gives a larger energy power of approximately 
$\om_p^{(1+2)} = 0.02 + 0.09 = 0.11$. 
This result should also be taken with some caution 
since the eigenfunctions are now oscillatory,
 $\psi(q) \propto (q^2)^{(i \nu_s)} $, and when folded in with 
the impact factors could lead to negative cross sections 
(in addition, it implies that the full power may not be seen until very high energies) !

The running coupling piece of the next-to-leading order kernel is 
potentially even more troublesome as pointed out 
by Armesto, Bartels and Braun \cite{abb}. 
The corresponding part in the Hamiltonian is a potential
piece proportional to $ \beta_0 \ln k^2/\mu^2$. 
Since this piece can 
take on any value, the energy spectrum of the Hamiltonian is unbounded 
from below which implies an arbitrary large power of energy is possible 
(there is no rightmost singularity in the $j$-plane and therefore no intercept).
Such extreme growth with energy is very worrying, however, as Armesto \etal 
point out, it comes with a non-perturbative damping factor, $\exp(-1/\as b)$,
which may lead to its suppression for the scattering of small objects.  
Assuming this to be the case the authors of \cite{abb}  
also reproduce the non-Regge type behaviour (a non-power-like behaviour in s)
arising from the running coupling found in \cite{km} (see also \cite{levin}), 
this behaviour also points to the restricted applicability of the whole formalism.

In practice we know that the unboundedness comes from the running coupling 
approaching the Landau pole. We know that there must be 
some regulation in this infra-red region which marks the transition between 
perturbative and non-perturbative QCD. 
The question which needs addressing is whether the way in which one 
implements this infra-red regularization \cite{lip,cc2} makes a critical difference 
to the energy dependence. If the answer is yes, it would appear
that non-perturbative physics is inextricably entangled with
perturbative physics, even in idealised onia-onia scattering, and that 
one loses all predictability in perturbative QCD at high enough energies.
If the answer is no, it may be possible to factorize all 
non-perturbative behaviour into boundary conditions and retain 
some predictability. This issue is intimately connected with the 
enhancement of the diffusion in $k^2$ caused by the running coupling. 
For a more detailed discussion of the effects of running coupling in the 
leading-order BFKL kernel see chapter (5) of \cite{fr}.

On a more optimistic note, DIS at moderate $x$ is well-understood 
and it would be very surprising if it was impossible to approach the small-$x$ 
region in a controlled way. A formalism exists to do this which 
incorporates information from BFKL dynamics into the
DGLAP \cite{dglap} formalism  by using resummed anomalous dimensions (see \cite{fl,cc,bv} 
and references therein).  
Thorne gave an interesting presentation on this issue, which included 
the possibility of an energy dependent coupling
constant in the small-$x$ region this may lead to enlargement of 
the region of applicability of the DGLAP formalism \cite{fst}.

One optimistic view of the large correction factor in equation (\ref{lc}) pointed 
out in the discussions is that what one is seeing is the 
first piece of some large all-orders effect which may be resummed. 
This all orders effect could be connected to coherence phenomena. 
Salam \cite{salam} gave a presentation in
which double logarithms in $k^2$ are resummed 
 and discussed the corresponding radical changes in the structure of the kernel 
(see \cite{fst} for more details).

As the summary above should indicate there are many issues that still need 
to be resolved in this fascinating area. It is clear from the discussions at Durham that 
the intense theoretical interest in this area is set to continue.

\ack{I'm happy to thank Jeff Forshaw for help in preparing my talk 
and the organisers for an excellent meeting.}


\section*{References}

\begin{thebibliography}{99}

\bibitem{bcox} B. Cox, these proceedings. 

\bibitem{wgb} K. Golec-Biernat and M. Wuesthoff, hep-ph/9807513. 

\bibitem{mkw} J. Jalilan-Marian, A. Kovner and H. Weigert, hep-ph/9709423.

\bibitem{h1diff} H1 Collab., Z. Phys. {\bf C76} (1997) 613.

\bibitem{h1dijet} H1 Collab., hep-ex/9808013.

\bibitem{whitmore} J. Whitmore, {\it Extracting diffractive parton distributions from HERA data and factorization tests}, 
Proc. of DIS98, April 1998, Brussels, to be published by World Scientific;
L. Alvero et al, hep-ph/9805268.


\bibitem{levin2} E. Gotsman, E. Levin and U. Maor, Phys. Lett. {\bf B438} 369.


\bibitem{dl} A. Donnachie and P. V. Landshoff, hep-ph/9806344


\bibitem{nhera} M. Arneodo, A. Bialas, M. W. Krasny. T. Sloan and M. Strikman 
{\it Nuclear beams at HERA}, Proc. of DESY 1995/1996 Workshop ``Future Physics at HERA'', ed  G. Ingelman, A de Roeck, R Klanner, 887.
             

\bibitem{ji} X.-D. Ji, \jpg{24}{1998}{1181}

\bibitem{critt} J. A. Crittenden, hep-ex/9806020. 

\bibitem{ik} D. Yu. Ivanov and R. Kirschner, hep-ph/9807324. 

\bibitem{fl} V. S. Fadin and L. N. Lipatov, Phys. Lett. {\bf B429}
(1998) 127.

\bibitem{cc} G. Camici and M. Ciafaloni, Phys. Lett. {\bf B412} (1997) 396.

\bibitem{bfkl} V. S. Fadin, E. A. Kuraev, L. N. Lipatov, \PL {\bf B60} (1975) 50; 
Ya. Ya. Balitsky, L. N. Lipatov, Sov. J. Nucl. Phys. {\bf 28} (1978) 822.

\bibitem{fr} J.R. Forshaw and D. A. Ross, 1997, {\it Quantum Chromodynmaics and the Pomeron}, (Cambridge: Cambridge University Press).

\bibitem{fadin} V. S. Fadin. {\it ``BFKL news''}, talk given at LISHEP98, February 1998, 
Rio de Janeiro, (hep-ph/9806482). 

\bibitem{dross} D. A. Ross, Phys. Lett. {\bf B431} (1998) 161.

\bibitem{abb} N. Armesto, J. Bartels and M. A. Braun, hep-ph/9808340 

\bibitem{km} Yu. V. Kovchegov and A. H. Mueller, preprint CU-TP-899
(hep-ph/9805208).

\bibitem{levin} E. M. Levin, preprint TAUP 2501-98 (hep-ph/9806228).

\bibitem{lip} L. N. Lipatov, Sov. Phys. JETP {\bf 63} (1986) 904.

\bibitem{cc2} G. Camici and M. Ciafaloni, Phys. Lett. {\bf B395} (1997) 118.

\bibitem{dglap} V. N. Gribov, L. N. Lipatov, Sov. J. Nucl. Phys. {\bf 15} (1972) 438;
L. N. Lipatov, Sov. J. Nucl. Phys. {\bf 20} (1975) 94;
G. Altarelli, G. Parisi, Nucl. Phys. {\bf B26} (1977) 298;
Yu. L. Dokshitzer, Sov. Phys. JETP {\bf 46} (1977) 641.

\bibitem{bv} J. Bl\"umlein and A. Vogt, Phys. Rev. {\bf D57} (1998) 1;
{\bf D58} (1998) 014020. J. Bl\"{u}mlein, V. Ravindram, W. L. van Neervan, A. Vogt, hep-ph/9806368.

\bibitem{fst} J. R. Forshaw, G. Salam and R. Thorne, these proceedings. 

\bibitem{salam} G. Salam, hep-ph/9806482 





\end{thebibliography}
\end{document}